\documentclass[twocolumn]{aastex631}
\usepackage{enumitem,natbib,amsmath}

\newcommand{\eV}{\mathrm{eV}}
\newcommand{\EeV}{\mathrm{EeV}}

\newcommand{\expo}{\mathcal{E}}
\newcommand{\num}{N}
\newcommand{\flux}{\Phi}
\newcommand{\ein}{\expo_\text{in}}
\newcommand{\eout}{\expo_\text{out}}
\newcommand{\etot}{\expo_\text{tot}}
\newcommand{\nin}{\num_\text{in}}
\newcommand{\nout}{\num_\text{out}}
\newcommand{\ntot}{\num_\text{tot}}
\newcommand{\nbg}{\num_\text{bg}}
\newcommand{\fin}{\flux_\text{in}}
\newcommand{\fout}{\flux_\text{out}}

\newcommand{\ZLM}{Z_\text{LM}}
\newcommand{\dd}{\mathrm{d}}

\definecolor{myred}{HTML}{ee0000}
\definecolor{mypurple}{HTML}{9370DB}
\definecolor{mygreen}{HTML}{26964f}
\definecolor{mysalmon}{HTML}{fa8072}

\submitjournal{\apj}
\shorttitle{The distribution of UHECRs along the SGP measured at the Auger Observatory}
\shortauthors{Pierre Auger Collaboration}

\begin{document}

\title{The distribution of ultra-high-energy cosmic rays along the supergalactic plane measured at the Pierre Auger Observatory}

\collaboration{0}{The Pierre Auger Collaboration}
\email{spokespersons@auger.org}

\begin{abstract}
Ultra-high-energy cosmic rays are known to be mainly of extragalactic origin, and their propagation is limited by energy losses, so their arrival directions are expected to correlate with the large-scale structure of the local Universe.
In this work, we investigate the possible presence of intermediate-scale excesses in the flux of the most energetic cosmic rays from the direction of the supergalactic plane region using events with energies above~20~EeV recorded with the surface detector array of the Pierre Auger Observatory up to 31 December 2022, with a total exposure of~$135{,}000\,\mathrm{km}^2\,\mathrm{sr}\,\mathrm{yr}$.
The strongest indication for an excess that we find, with a post-trial significance of~$3.1\sigma$, is in the Centaurus region, as in our previous reports, and it extends down to lower energies than previously studied.
We do not find any strong hints of excesses from any other region of the supergalactic plane at the same angular scale.
In particular, our results do not confirm the reports by the Telescope Array collaboration of excesses from two regions in the Northern Hemisphere at the edge of the field of view of the Pierre Auger Observatory. With a comparable integrated exposure over these regions, our results there are in good agreement with the expectations from an isotropic distribution.
\end{abstract}


\section{Introduction} 
\label{sec:intro}

The flux of ultra-high-energy cosmic rays (UHECRs), atomic nuclei mainly of extragalactic origin reaching the Earth with energies~$E \ge 1\,\EeV = 10^{18}\,\eV \approx 0.16\,\mathrm{J}$ each, is remarkably close to being the same from all directions in the sky, with the exception of a dipole moment in the celestial distribution of cosmic rays with~$E \ge 8\,\EeV$ \citep{dipole2017,dipole2018,dipole2020} towards a direction $\sim 115\degr$~away from the Galactic Center, with an amplitude of around~6\% at~$10\,\EeV$ and growing roughly linearly with energy.  No anisotropies on intermediate or smaller angular scales have been conclusively discovered yet in data collected at either the Pierre Auger Observatory or the Telescope Array (TA), the two largest cosmic-ray detector arrays in the world (covering $3000\,\mathrm{km}^2$ and $700\,\mathrm{km}^2$ respectively), located in the Southern and Northern Hemisphere (latitudes~$-35\fdg2$ and~$+39\fdg3$), respectively.
On the other hand, a few indications with statistical significances ranging from~$3.0\sigma$ to~$4.6\sigma$ of such anisotropies in the flux of cosmic rays with more than a few tens of EeV have been reported.
An excess of events in data from the Pierre Auger Observatory from a circular region on the celestial sphere centered on the Centaurus~A (Cen~A) radio galaxy, first reported in \citet{Auger_APh2010}, has reached a post-trial significance of~$4.0\sigma$ \citep{Auger_ICRC2023}.
A correlation with the positions of nearby starburst galaxies first reported in \citet{Auger_ApJL2018}, to which the main contributor is the NGC4945 galaxy in the aforementioned Cen~A region, has reached $3.8\sigma$ post-trial as of the last update \citep{Auger_ICRC2023}.
An analogous study combining data from both the Pierre Auger Observatory and the Telescope Array has reached $4.6\sigma$ post-trial \citep{joint_ICRC2023}.
Finally, the so-called ``TA hotspot'' at equatorial coordinates~$(\alpha,\delta)\approx(145\degr,+40\degr)$ \citep{TA_ApJL2014} and a new excess at~$(\alpha,\delta)\approx(20\degr,+35\degr)$ \citep{TA_arXiv2021} in TA data have post-trial significances around~$3\sigma$ as of their last update \citep{TA_ICRC2023}.
All these regions where indications of excesses have been reported intersect the supergalactic plane (SGP), a great circle in the sky along which extragalactic matter within~$\mathcal{O}(10^2\,\mathrm{Mpc})$ tends to be concentrated.  The Local Sheet, a structure comprising nearly all bright galaxies within~$6\,\mathrm{Mpc}$ \citep{McCall}, is also remarkably aligned with the SGP.  Hence, a concentration of the flux of the highest-energy cosmic rays along this plane would not be completely unexpected, given that propagation lengths at the highest energies are limited to a few hundred Mpc---or even less, in the case of intermediate-mass nuclei \citep{propagation}. On the other hand, UHECRs can undergo substantial deflections by Galactic and possibly intergalactic magnetic fields \citep{cohGMF,turGMF,IGMF}, preventing a one-to-one interpretation of arrival directions in terms of source positions.

Here, we use the intermediate angular scale of the aforementioned excess from the region reported in data from the Pierre Auger Observatory.  As of the last update \citep{Auger_ICRC2023}, the maximum local Li--Ma significance for a circular window was achieved with an energy threshold of~$E_{\min} = 38\,\EeV$ and a window radius of~$\Psi=27\degr$, whether the center of the window was constrained to be the position of Cen~A or also scanned to avoid any assumption on the possible source location. In this work, we study whether other excesses with similar characteristics are present in different regions along the SGP, and/or at lower energies than previously studied.
A search for excesses of events in bands centered around the SGP found no statistically significant result \citep[$p=0.13$~post-trial,][section~3.3]{Auger_ApJ2022}, but a band in latitude may not capture an excess concentrated in a limited range of supergalactic longitude, hence in this work we consider top-hat windows (i.e., disk search regions) intersecting the SGP instead.

\section{The dataset}
\label{sec:data}

We use the same dataset used in our last update on arrival directions \citep{Auger_ICRC2023} for searches for medium-scale anisotropies, namely events recorded using the surface detector (SD) array of the Pierre Auger Observatory in the years from 2004 to 2022 inclusive.  We only consider events with energies~$E \ge 20\,\EeV$, as in \citet{Auger_ApJL2018}.
This is the same as the published dataset of \citet{Auger_ApJ2022} with the addition of the events detected in the years 2021 and 2022 and of events with energies~$20\,\EeV \le E < 32\,\EeV$ over the entire time period.  As regards the last two years, only events detected by the parts of the array that had not yet undergone the AugerPrime upgrade \citep{Auger_Prime} are used.
As in \citet{Auger_ApJ2022}, we use all ``vertical'' events (with zenith angles~$\theta < 60\degr$) in which the SD station with the largest signal is surrounded by at least four active stations and the reconstructed shower core is within an isosceles triangle of active stations, and all ``inclined'' events (with~$60\degr \le \theta < 80\degr$) in which the station closest to the reconstructed core position is surrounded by at least five active stations.  
The energies of these events are reconstructed with a total systematic uncertainty~$\sim 14\%$ and resolution~$\sim 7\%$, and their arrival directions with a resolution~$< 1\degr$, for both vertical and inclined events.
The total exposure of this dataset is $135{,}000\,\mathrm{km}^2\,\mathrm{sr}\,\mathrm{yr}$.

As in \citet{Auger_ApJ2022,Auger_ICRC2023}, the exposures to vertical and inclined events are rescaled so as to be proportional to the number of events in each zenith angle range (respectively~6896 and~1936 above~$20\,\EeV$).  
We checked that the ratio between the number of inclined and vertical events, $0.281\pm0.007$, is within statistical uncertainties of the expectation~$0.278$ \citep[section~2 and appendix~A]{Auger_ApJ2022}.
The rescaling of exposures ensures that our analysis is very robust to any systematic effects affecting vertical and inclined events separately: we find that even artificially multiplying or dividing all inclined energies by a factor~$1.25$ before the rescaling
would affect the resulting
flux in the circular regions of the Southern sky listed in \autoref{tab:results} by less than the statistical uncertainties.
By combining both zenith angle ranges ($0\degr \le \theta < 80\degr$), the field of view (FoV) of the SD array covers all declinations~$-90\degr \le \delta < +44\fdg8$.

\section{Analysis method}

In this work, for each of six different energy thresholds, $E_{\min} = 20, 25, 32, 40, 50, 63\,\EeV$
(i.e., $10^{19.3, 19.4, \ldots, 19.8}\,\eV$ rounded to the nearest~EeV), we consider all circular windows with radius~$\Psi = 27\degr$ (the maximum-significance radius in \citealt{Auger_ICRC2023}) centered on the positions on a HEALPix\footnote{\url{https://healpix.sourceforge.io/}} grid \citep{HEALPix} with~$N_\text{side}=2^{6}$ (resolution~$\approx 0\fdg9$) simultaneously meeting two criteria: first, we require that the SGP intersect the window, i.e., that the supergalactic latitude~$B$ of the window center satisfy~$-\Psi \le B \le \Psi$; and second, as in our previous works, in order to have reasonably large statistics we require that the center of the window be inside the FoV of the Observatory, i.e., that the declination of the window center satisfy~$\delta < +44\fdg8$.\footnote{Note that this is slightly more conservative than the recommendation by \citet{LiMa} that~$\nin\gtrsim10$ and~$\nout\gtrsim10$ when using the lowest of the energy thresholds we use here but slightly less conservative using the highest thresholds, i.e., some of the windows with centers closest to the edge of the FoV have~$\nin \lesssim 10$ when using the highest thresholds.}
For each such window, we counted the numbers~$\nin,\nout$ of events in our dataset with~$E \ge E_{\min}$
respectively
inside the window and in the rest of the FoV,
and computed the exposures $\ein, \eout$ by numerically integrating the expression in \citet[][section~2]{Sommers}.
From these, we computed the background number of events~$\nbg$ as~$\nout\ein/\eout$, and the flux ratio~$\fin/\fout$ as~$\nin/\nbg$ (see below).

\subsection{Binomial probability, likelihood and upper limit}
\label{sec:lik}

For a given value of the ratio~$\fin/\fout$ between the flux inside the window and that in the rest of the FoV (the isotropic null hypothesis being~$\fin/\fout = 1$) and total number~$\ntot=\nin+\nout$ of events above the energy threshold,
the probability to observe exactly~$\nin$ events inside the window is \begin{equation}
        P\left(\nin \middle| \ntot, \tfrac\fin\fout \right)
    = \binom\ntot\nin p^{\nin} (1-p)^{\ntot-\nin},\label{eq:binomial}
\end{equation} where \begin{equation}
    p = \frac{\fin\ein}{\fin\ein + \fout\eout}
\end{equation} is the probability for each event to fall within the window.
This probability as a function of~$\fin/\fout$ for a fixed~$\nin,\nout$ defines a likelihood function, \begin{equation}
    L(\fin/\fout) = P\left(\nin \middle| \ntot, \fin/\fout \right),\label{eq:likelihood}
\end{equation}
which achieves its maximum at~$\fin/\fout = \frac{\nin/\ein}{\nout/\eout} = \nin/\nbg$.

If we define the deviance (generalized~$\chi^2$, here with one degree of freedom) as \begin{align}
    D(\fin/\fout)
    &= -2\ln\frac{L(\fin/\fout)}{\max_{\fin/\fout}L(\fin/\fout)} \nonumber\\
    &= -2\ln\frac{L(\fin/\fout)}{L(\nin/\nbg)},\label{eq:deviance}
\end{align} then $\pm\sqrt{D(\fin/\fout)}$~is the number of standard deviations at which the dataset disfavors a given value of~$\fin/\fout$ with respect to the value~$\nin/\nbg$; in particular, $\pm\sqrt{D(\fin/\fout=1)}$~equals the local Li--Ma significance~$\ZLM$ \citep{LiMa}.\footnote{The sign is~$+$ or~$-$ depending on whether $\fin/\fout$ is larger or smaller than the maximum-likelihood value~$\nin/\nbg$.}  The statistical uncertainties in~$\fin/\fout$ we report in the tables are the $\pm1\sigma$ intervals defined this way.

Finally, we define the frequentist 99\%~confidence level upper limit to~$\fin/\fout$ as the $\fin/\fout$~value such that \begin{equation}
    \sum_{n=\nin+1}^{\ntot} P\left(n \middle| \ntot, \fin/\fout \right) = 0.01; \label{eq:upper}
\end{equation} in the cases we report, this agrees with the value such that $\sqrt{D(\fin/\fout)} = 2.33$ to within a few percent.

\section{Results}

\begin{figure*}
    \plotone{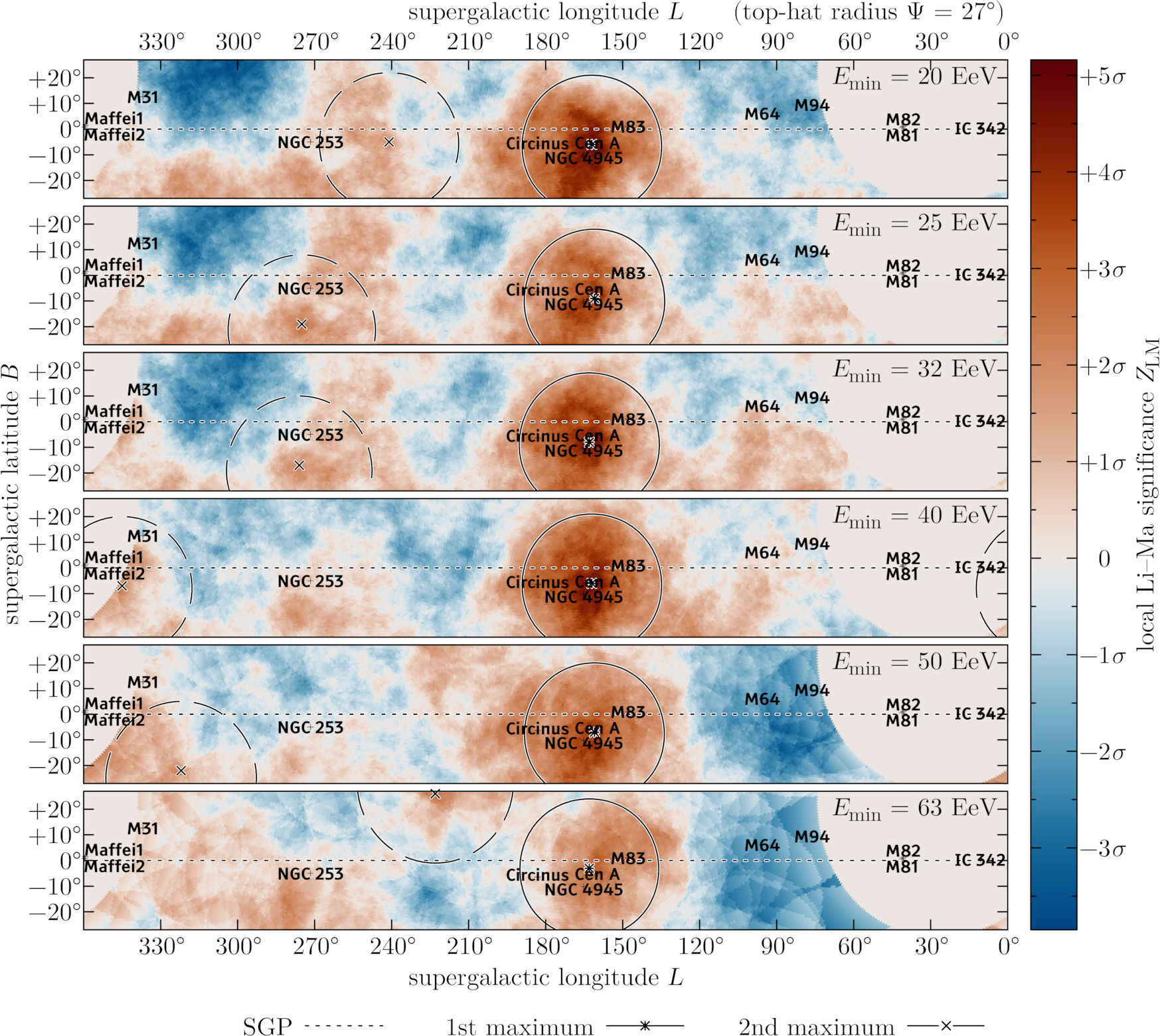}
    \caption{Local Li--Ma significance~$\ZLM$ of excesses over the isotropic expectation as a function of the window center position.  The~$\ZLM$ in windows whose center lies outside the FoV of the Observatory was not computed (shown as the gray disk wrapping around the left and right edges of each panel; see also \autoref{fig:regions}).
    In each panel, the energy threshold used is written in the upper right corner.
    The solid circle is the window position with the highest~$\ZLM$ in the whole strip; the dashed one is that with the highest~$\ZLM$ excluding those overlapping with the solid one.
    Labels indicate the position of Council of Giants galaxies \citep{McCall} for reference only; they are not taken into account in the analysis in any way.
    \label{fig:lima}}
\end{figure*}

\begin{figure*}
    \plotone{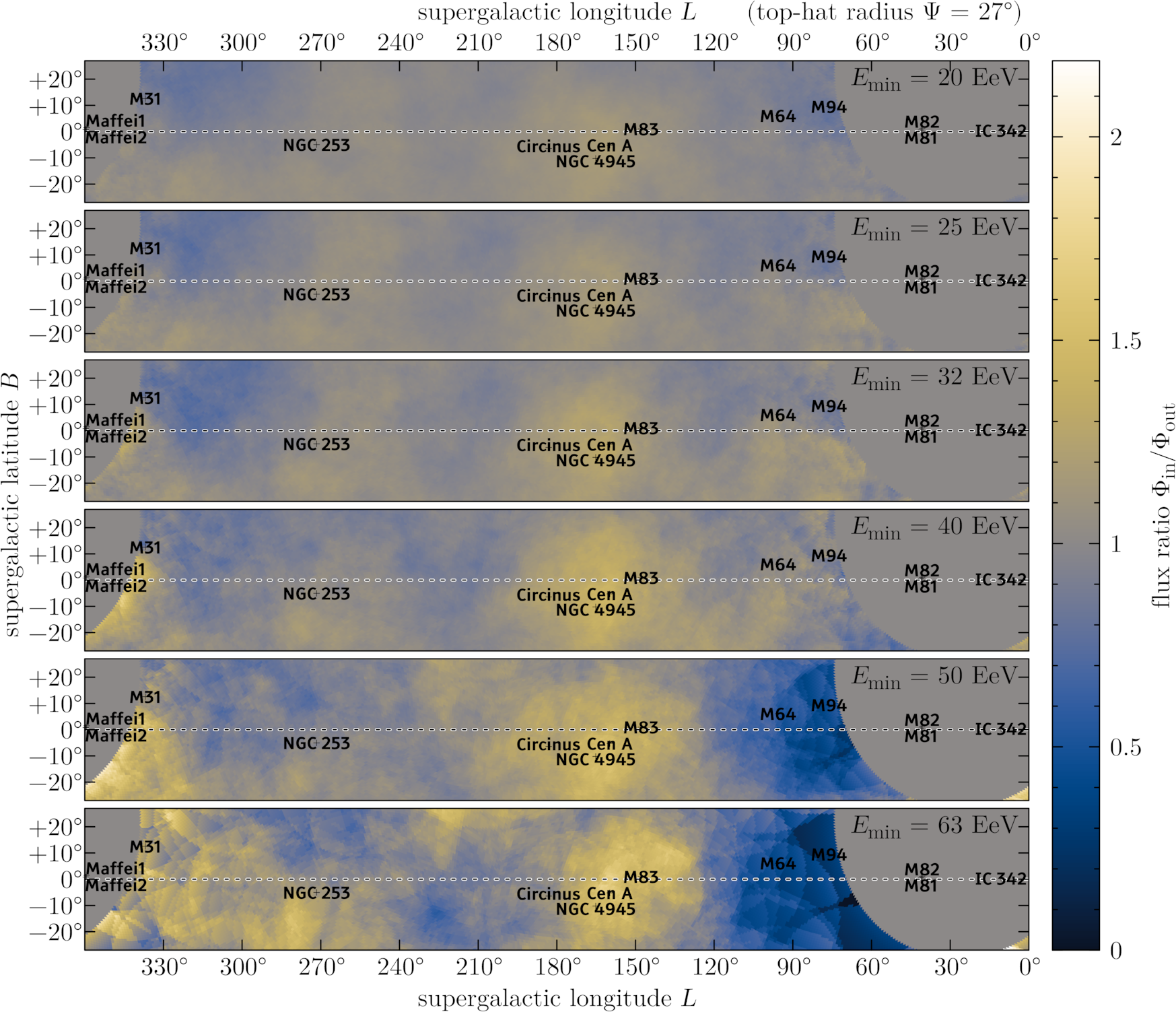}\\
    \caption{The maximum-likelihood value of the ratio~$\fin/\fout$, i.e., $\nin/\nbg$, as a function of the window center position
    \label{fig:ratio}}
\end{figure*}

The local Li--Ma significance~$\ZLM$ as a function of the position of the window center in supergalactic coordinates~$(L,B)$ is shown in \autoref{fig:lima},
and the information about the window with the highest~$\ZLM$ for each~$E_{\min}$ is provided in \autoref{tab:results}.
\begin{table*}
    \centering
    \caption{Information about the maximum-significance excesses found along the SGP\label{tab:results}}
    \setlength{\tabcolsep}{3.4pt}
    \begin{tabular}{CC|CCCCCCCC|CCCCCCCC|}
        \multicolumn{2}{c}{\strut} & \multicolumn{8}{c}{1st maximum} & \multicolumn{8}{c}{2nd maximum} \\
        E_{\min} & \ntot & 
        L & B & \tfrac\ein\etot & \nbg & \nin & \tfrac\fin\fout & \ZLM & ^{99\%}_\mathrm{U.L.} & 
        L & B & \tfrac\ein\etot & \nbg & \nin & \tfrac\fin\fout & \ZLM & ^{99\%}_\mathrm{U.L.} \\[\smallskipamount]
        \tableline
        20~\EeV & 8832 & 162\degr & -6\degr & 9.56\% & 829. & 990 & 1.19^{+0.04}_{-0.04} & +5.2\sigma & 1.29 & 241\degr & \phn{-}5\degr & 10.27\% & 900. & 971 & 1.08^{+0.04}_{-0.04} & +2.2\sigma & 1.17 \\
25~\EeV & 5380 & 161\degr & -9\degr & 9.56\% & 504. & 608 & 1.21^{+0.05}_{-0.05} & +4.2\sigma & 1.33 & 275\degr & -19\degr & \phn8.00\% & 426. & 482 & 1.13^{+0.05}_{-0.05} & +2.6\sigma & 1.26 \\
32~\EeV & 2936 & 163\degr & -8\degr & 9.68\% & 276. & 363 & 1.32^{+0.08}_{-0.07} & +4.7\sigma & 1.50 & 276\degr & -17\degr & \phn7.89\% & 229. & 264 & 1.15^{+0.08}_{-0.07} & +2.2\sigma & 1.34 \\
40~\EeV & 1533 & 162\degr & -6\degr & 9.56\% & 140. & 208 & 1.49^{+0.11}_{-0.11} & +5.1\sigma & 1.77 & 345\degr & \phn{-}7\degr & \phn1.00\% & 15.2 & \phn26 & 1.71^{+0.36}_{-0.32} & +2.5\sigma & 2.68 \\
50~\EeV & \phn713 & 161\degr & -7\degr & 9.56\% & 64.4 & 103 & 1.60^{+0.18}_{-0.16} & +4.2\sigma & 2.05 & 322\degr & -22\degr & \phn3.69\% & 25.9 & \phn39 & 1.51^{+0.26}_{-0.23} & +2.4\sigma & 2.20 \\
63~\EeV & \phn295 & 163\degr & -3\degr & 9.56\% & 26.3 & \phn46 & 1.75^{+0.30}_{-0.26} & +3.3\sigma & 2.54 & 223\degr & +26\degr & \phn9.56\% & 26.7 & \phn42 & 1.57^{+0.28}_{-0.25} & +2.6\sigma & 2.31 \\

        \tableline
    \end{tabular}
\end{table*}
We also search for the highest~$\ZLM$ among windows which do not overlap with the global maximum one (distance between centers~$> 2\Psi$).
In \autoref{fig:ratio},
we show the flux ratio~$\fin/\fout$ computed as~$\nin/\nbg$ as a function of the position of the window center.

\subsection{Indication of an excess in the Centaurus region}
As shown in the left part of \autoref{tab:results} and by the solid circles in \autoref{fig:lima}, with all energy thresholds the most significant excess is the previously reported one in the Centaurus region. Its position is remarkably stable at least over a range of energy thresholds spanning half an order of magnitude (and of cumulative UHECR flux values spanning one and a half orders of magnitude), with no discernible change in the maximum-significance window center.
On the other hand, the strength~$\fin/\fout$ of the excess does grow with the energy threshold, implying that the particles making up the excess have a different energy spectrum than the background, with a slower decrease with energy.
By studying the number of events in this region in separate energy bins  (see \autoref{app:bins} for details), 
we find that the excess has a spectral index~$\gamma=2.6\pm0.3$. For comparison, the overall spectrum in our FoV \citep[with stricter quality cuts and a different reconstruction]{spectrum2020} has $\gamma=3.05\pm0.05\pm0.10$ below~$(46\pm3\pm6)\,\EeV$ and $\gamma=5.1\pm0.3\pm0.1$ above, where the first uncertainty is statistical and the second is systematic.

The local significance of~$+5.2\sigma$ we find in the Centaurus region using the lowest energy threshold is exceeded for at least one of the window positions and energy thresholds in~912 out of~$10^6$ isotropic simulations, corresponding to a $3.1\sigma$ post-trial significance. 

\subsection{Study of Telescope Array reported excess regions}
\begin{figure*}
    \plotone{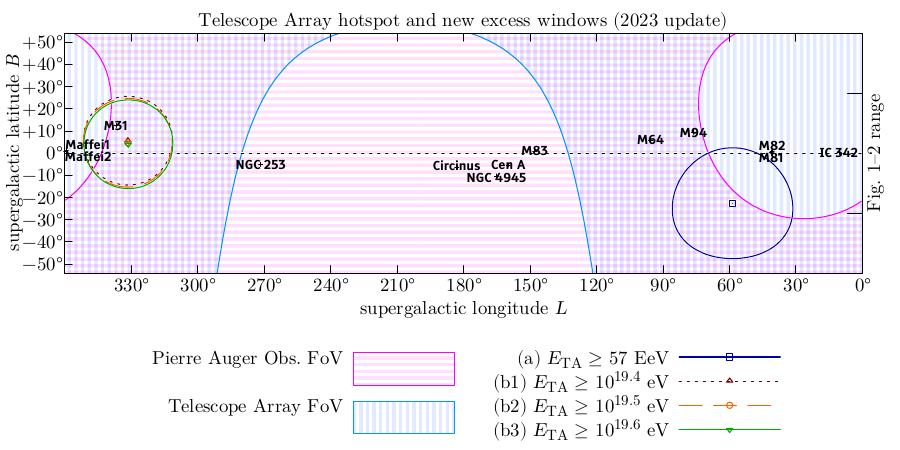}
    \caption{The windows in which the TA collaboration reported excesses of events, as of their latest update \citep{TA_ICRC2023}, compared to the FoV of the Pierre Auger Observatory and of the Telescope Array\label{fig:regions}}
\end{figure*}

As shown in the right part of \autoref{tab:results} and by the dashed circles in \autoref{fig:lima},
the local significances
of excesses in windows not overlapping with the maximum-significance one are below $2.7\sigma$
for all the energy thresholds we tested.  As shown in \autoref{app:ul}, this sets stringent upper limits on the flux, except very close to the edge of our FoV.
The non-observation of other excesses at this angular scale appears to contradict the reports by
the TA collaboration of an excess of cosmic rays with energies~$E \ge 57\,\EeV$ from a circular window (hereafter ``TA hotspot'') around~$(\alpha,\delta) \approx (145\degr, +40\degr)$ \citep{TA_ApJL2014}
and later of a weaker excess of events with~$E \ge 10^{19.4, 19.5, 19.6}\,\eV$ from a window around~$(\alpha,\delta)\approx(20\degr,+35\degr)$ \citep{TA_arXiv2021}, both shown in \autoref{fig:regions}. 
These reports have global statistical significances~$\sim 3\sigma$ so far \citep[as of][]{TA_ICRC2023}, but nevertheless they have already been met with considerable interest in the community and spurred several attempts at phenomenological interpretations \citep[e.g.,][]{Neronov,Plotko,Anchordoqui}.

\begin{figure}
    \plotone{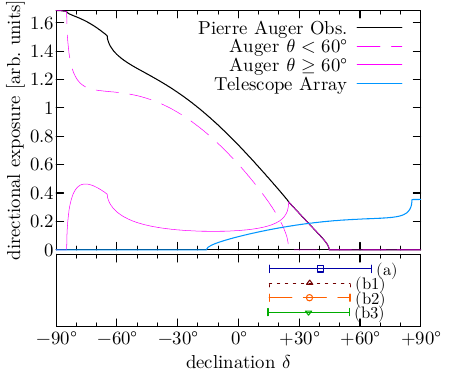}
    \caption{The directional exposure of the Pierre Auger Observatory and the Telescope Array as a function of declination, compared to the declinations of the windows shown in \autoref{fig:regions} (horizontal bars; note that the bar lengths denote the sizes of the windows, not the uncertainties in their position).
    \label{fig:exposures}}
\end{figure}
Since these windows intersect the SGP and their centers are inside the FoV of the Auger Observatory (see also \autoref{fig:exposures}), they are among the range of window centers we considered. As shown in \autoref{fig:lima}, we do not find any excesses at these positions when using comparable energy thresholds.

To find out what we could have expected to observe in our data given those reports from TA, 
after correcting the energy thresholds for the known mismatch between the energy scales of the two observatories \citep[eq.~(1)]{joint_ICRC2023},
we computed the distribution of the number~$\nin$ of events in our dataset expected in each of these windows based on (i)~isotropy ($\fin/\fout=1$), (ii)~the TA value of~$\fin/\fout$ that can be computed from their numbers of events~$\nin,\ntot$ as reported in their last update \citep{TA_ICRC2023}, or (iii)~the marginal distribution of $\fin/\fout$ over TA statistical uncertainties.  
\begin{figure} 
    \plotone{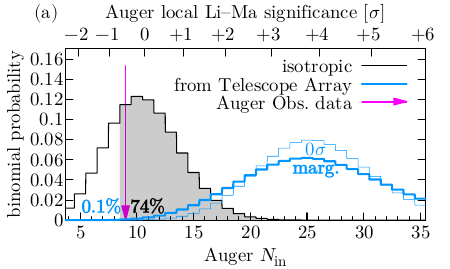}
    \caption{Binomial probability that $\nin$~events would be observed in our dataset in the first (a) of the windows reported by TA and shown in \autoref{fig:regions}. The thin blue histogram assumes that the value of the flux ratio~$\fin/\fout$ is exactly the one reported by TA, whereas the thick one is the marginal distribution of~$\fin/\fout$ over TA statistical uncertainties.  The complete figure set (4 images) is available in the online journal.\label{fig:binomial}}
    \figsetstart
        \figsetnum{5}
        \figsettitle{Binomial probability that $\nin$~events would be observed in our dataset in each of the windows reported by TA and shown in \autoref{fig:regions}. The thin blue histogram assumes that the value of the flux ratio~$\fin/\fout$ is exactly the one reported by TA, whereas the thick one is the marginal distribution of~$\fin/\fout$ over TA statistical uncertainties.}
        \figsetgrpstart
            \figsetgrpnum{5.1}
            \figsetgrptitle{(a)}
            \figsetplot{binom-a.pdf}
            \figsetgrpnote{First window (TA hotspot, $E_\min = 57\,\EeV_\text{TA}$)}
        \figsetgrpend
        \figsetgrpstart
            \figsetgrpnum{5.2}
            \figsetgrptitle{(b1)}
            \figsetplot{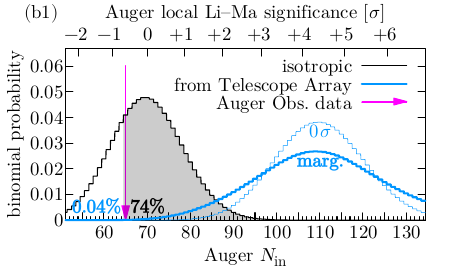}
            \figsetgrpnote{Second window (PPSC, $E_\min = 10^{19.4}\,\eV_\text{TA}$)}
        \figsetgrpend
        \figsetgrpstart
            \figsetgrpnum{5.3}
            \figsetgrptitle{(b2)}
            \figsetplot{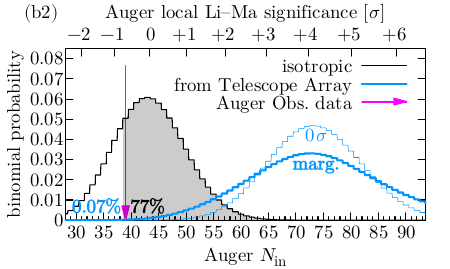}
            \figsetgrpnote{Third window (PPSC, $E_\min = 10^{19.5}\,\eV_\text{TA}$)}
        \figsetgrpend
        \figsetgrpstart
            \figsetgrpnum{5.4}
            \figsetgrptitle{(b3)}
            \figsetplot{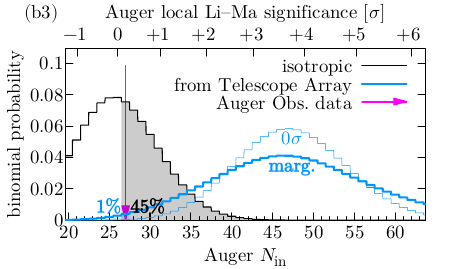}
            \figsetgrpnote{Fourth window (PPSC, $E_\min = 10^{19.6}\,\eV_\text{TA}$)}
        \figsetgrpend
    \figsetend
\end{figure}
As we show in \autoref{fig:binomial}, 
in each case we find that based on the TA result we would expect on average a local Li--Ma significance in our data of the order of $4\sigma$, comparable to the TA value---since the integrated exposures we have accumulated in these windows are comparable to TA ones, as shown by the $\nbg$~values\footnote{%
    We cannot compute the absolute integral exposure of TA within each window (in~$\mathrm{km}^2\,\mathrm{sr}\,\mathrm{yr}$) to directly compare it with ours, as \citet{TA_ICRC2023} did not report the total exposure of the dataset.  (\Citealt{joint_ICRC2023} did, but a different TA dataset with stricter selection criteria was used there.)%
} in \autoref{tab:ta}. 
\begin{table*}
    \centering
    \caption{The excesses reported by TA in the windows shown in \autoref{fig:regions}, as of their latest update \citep{TA_ICRC2023}, and the corresponding results in our data.
    The $E_{\min}$~values are converted from the TA energy scale to ours using \citet[eq.~(1)]{joint_ICRC2023}.
    Some of the TA values of $\nbg$, $\fin/\fout$ and/or $\ZLM$ shown here differ by up to a few percent from those reported in \citet{TA_ICRC2023}, 
    presumably because in that work $\ein/\etot$~was estimated from a Monte Carlo simulation with 100,000 events (of which $\mathcal{O}(10^4)$ within the window, hence with fluctuations~$\sim 1\%$ in~$\ein$), whereas here we computed it by numerically integrating the expression in \citet[][section~2]{Sommers} over a HEALPix grid with~$N_\text{side}=2^{10}$ (resolution~$\approx 0\fdg06$).  
    For the TA results, we computed the frequentist 99\%~CL lower limit to~$\fin/\fout$ defined analogously to~\eqref{eq:upper} by~$\sum_{n=0}^{\nin-1} P\left(n \middle| \ntot, \fin/\fout \right) = 0.01$.
    Note that the TA post-trial significances were computed under the assumption that only excesses near the center of a presumed emitting structure (the Perseus--Pisces Supercluster) had been searched for. 
    \label{tab:ta}}
    \setlength{\tabcolsep}{2.8pt}
    \begin{tabular}{r|CCCCCCCCC|CCCCCCCC|}
        \multicolumn{1}{c}{\strut} & \multicolumn{9}{c}{Telescope Array \citep{TA_ICRC2023}} & \multicolumn{8}{c}{Pierre Auger Observatory (this work)} \\
        \multicolumn{1}{c|}{\strut} &
        E_{\min} & \ntot & \tfrac\ein\etot & \nbg & \nin & \tfrac\fin\fout & \ZLM & ^{99\%}_\mathrm{L.L.} & ^\text{post-}_\text{trial} &
        E_{\min} & \ntot & \tfrac\ein\etot & \nbg & \nin & \tfrac\fin\fout & \ZLM & ^{99\%}_\mathrm{U.L.} \\[\smallskipamount]
        \tableline
        (a) & 57~\EeV & \phn216 & 9.47\% & 18.0 & \phn44 & 2.44^{+0.44}_{-0.39} & +4.8\sigma & 1.60 & 2.8\sigma & 44.6~\EeV & 1074 & 1.00\% & 10.7 & \phn9 & 0.84^{+0.31}_{-0.25} & -0.5\sigma & 1.76 \\
(b1) & 10^{19.4}\,\eV & 1125 & 5.88\% & 64.0 & 101 & 1.58^{+0.17}_{-0.16} & +4.1\sigma & 1.22 & 3.3\sigma & 20.5~\EeV & 8374 & 0.84\% & 70.1 & 65 & 0.93^{+0.12}_{-0.11} & -0.6\sigma & 1.23 \\
(b2) & 10^{19.5}\,\eV & \phn728 & 5.87\% & 41.1 & \phn70 & 1.70^{+0.22}_{-0.20} & +4.0\sigma & 1.25 & 3.2\sigma & 25.5~\EeV & 5156 & 0.84\% & 43.5 & 39 & 0.90^{+0.15}_{-0.14} & -0.7\sigma & 1.29 \\
(b3) & 10^{19.6}\,\eV & \phn441 & 5.84\% & 24.6 & \phn45 & 1.83^{+0.31}_{-0.27} & +3.6\sigma & 1.23 & 3.0\sigma & 31.7~\EeV & 2990 & 0.87\% & 26.0 & 27 & 1.04^{+0.21}_{-0.19} & +0.2\sigma & 1.61 \\

        \tableline
    \end{tabular}
\end{table*}
Instead, what we actually obtain is always~$-0.7\sigma \lesssim \ZLM < +0.2\sigma$, in excellent agreement with the isotropic null hypothesis. In all cases, there exist possible values of~$\fin/\fout$ which would be compatible with both the 99\%~CL lower limit from TA data and the 99\%~CL upper limit from our data, e.g., $1.60 < \fin/\fout < 1.76$ in~(a).

\begin{figure*}
    \plotone{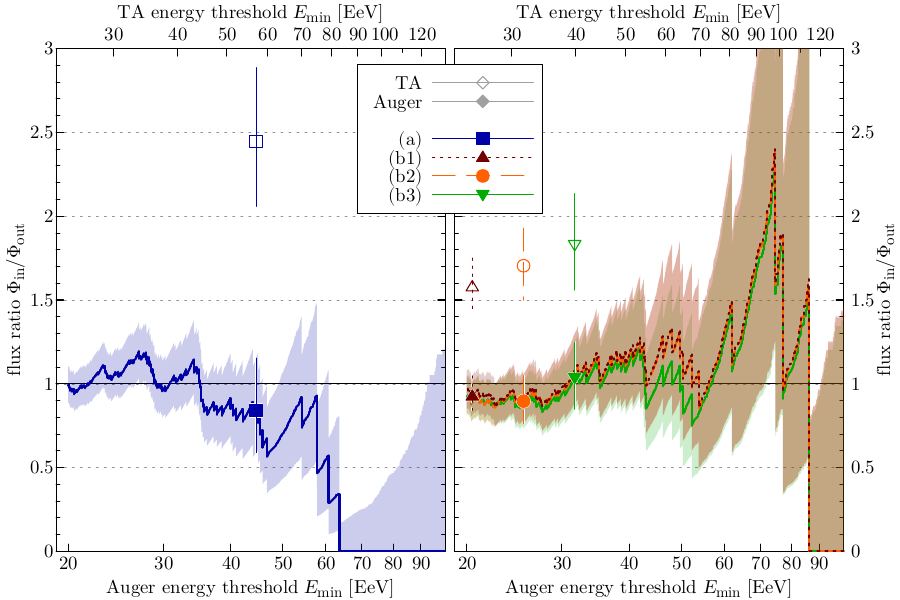}
    \caption{Flux ratio in the windows considered in this work (a, b1, b2, b3) computed from our data with all possible energy thresholds. The shaded bands show the $\pm1\sigma$~interval for each threshold.
    The filled markers indicate the values at the thresholds computed via \citet[eq.~(1)]{joint_ICRC2023}; the uncertainties on the energy cross-calibration are comparable to the horizontal size of the markers.  The results reported in \cite{TA_ICRC2023} are also shown as empty markers for comparison.
    \label{fig:energy_scan}}
\end{figure*}

\Citet{TA_coldspot} reported a deficit of events in the same part of the sky as the TA hotspot and at immediately lower energies, so that in a search for anisotropies with an energy threshold lower than that of the hotspot they would partially cancel each other out.  To take into account the possibility that \citet[eq.~(1)]{joint_ICRC2023} under- or overestimates the energy on the Pierre Auger Observatory scale corresponding to a given energy on the Telescope Array because of statistical and systematic uncertainties in the fit, we also computed $\fin/\fout$~and $\ZLM$~values with different energy thresholds, finding that no other choice of threshold yields significances comparable to what we would expect based on the TA results, either (\autoref{fig:energy_scan}).

A limitation of this study is that the likelihood in \autoref{eq:likelihood} implicitly assumes a constant UHECR flux~$\fin$ inside the window being considered and a constant flux~$\fout$ outside.  Whereas the directional exposure of the Telescope Array is roughly uniform within the windows shown in \autoref{fig:regions}, that of the Pierre Auger Observatory steeply decreases with increasing declination (and even vanishes in part of the windows), as shown in \autoref{fig:exposures},
so a flux excess more concentrated in the northern than in the southern part of a window would on average be underestimated when using data from the Auger Observatory.  
On the other hand, it should be noted that the TA window positions were not fixed a priori, but chosen in order to maximize the statistical significance of the excesses using TA data.  This indicates that the excess is located roughly equally in the northern and southern parts of the window: if more of the excess were in the northern than in the southern part of the window, the significance would have been maximized by a different window further north, and vice versa.
Similar considerations could apply to a declination dependence of the flux outside the window being considered, but in \citet{spectrum2020} we found that the flux of UHECRs does not appreciably vary with declination within our FoV other than the dipole mentioned in \autoref{sec:intro}. Furthermore, the declination dependence in the TA FoV claimed in \citet{TA_decldep,TA_arXiv2024}, if anything, would make the Telescope Array overestimate and the Auger Observatory underestimate~$\fout$, going in the opposite direction than what would explain away the difference between the $\fin/\fout$~values from the two datasets.

\section{Discussion and conclusions}
We have confirmed our previous finding \citep[with $4.0\sigma$ post-trial there]{Auger_ICRC2023} that the statistically most significant excess of UHECRs along the SGP is from the Centaurus region, though still not at the discovery level with the current statistics (post-trial significance $3.1\sigma$ for a fixed search radius in this work), and we have further found that this excess extends to lower energies than previously studied (down to~$20~\EeV$), with no appreciable dependence of its position on the energy threshold chosen. 
In case future experiments with more statistics confirm this excess, one possible explanation for the lack of energy dependence of its position (other than the absence of sizable coherent magnetic deflections) could be an approximately constant magnetic rigidity~$R=E/Z$ of the particles in it, i.e., an increasingly heavy mass composition such that their atomic numbers~$Z$ are proportional to their energy.  

On the other hand, no statistically significant excesses were found in the regions where TA reported excesses of events, despite comparable integral exposures in those regions.
It will be interesting to see whether the  AugerPrime~\citep{Auger_Prime} and TA$\times$4~\citep{TAx4} upgraded detectors or future observatories such as GRAND~\citep{GRAND}, POEMMA~\citep{POEMMA} or GCOS~\citep{GCOS} will confirm or rule out the indications for excesses reported by current experiments, and/or detect other anisotropies too weak to be noticed with the number of events gathered so far by current observatories.
If any excesses are confirmed, event-by-event mass information from upgraded detectors \citep{Auger_Prime} and/or machine learning techniques \citep{ML_muons,ML_depth,TA_BDT} will help us elucidate their origin in the future by examining whether and how the mass composition in such regions differs from that in the rest of the sky
and the energy dependence of any such differences.

\section*{Acknowledgments}

\begin{sloppypar}
The successful installation, commissioning, and operation of the Pierre
Auger Observatory would not have been possible without the strong
commitment and effort from the technical and administrative staff in
Malarg\"ue. We are very grateful to the following agencies and
organizations for financial support:
\end{sloppypar}

\begin{sloppypar}
Argentina -- Comisi\'on Nacional de Energ\'\i{}a At\'omica; Agencia Nacional de
Promoci\'on Cient\'\i{}fica y Tecnol\'ogica (ANPCyT); Consejo Nacional de
Investigaciones Cient\'\i{}ficas y T\'ecnicas (CONICET); Gobierno de la
Provincia de Mendoza; Municipalidad de Malarg\"ue; NDM Holdings and Valle
Las Le\~nas; in gratitude for their continuing cooperation over land
access; Australia -- the Australian Research Council; Belgium -- Fonds
de la Recherche Scientifique (FNRS); Research Foundation Flanders (FWO),
Marie Curie Action of the European Union Grant No.~101107047; Brazil --
Conselho Nacional de Desenvolvimento Cient\'\i{}fico e Tecnol\'ogico (CNPq);
Financiadora de Estudos e Projetos (FINEP); Funda\c{c}\~ao de Amparo \`a
Pesquisa do Estado de Rio de Janeiro (FAPERJ); S\~ao Paulo Research
Foundation (FAPESP) Grants No.~2019/10151-2, No.~2010/07359-6 and
No.~1999/05404-3; Minist\'erio da Ci\^encia, Tecnologia, Inova\c{c}\~oes e
Comunica\c{c}\~oes (MCTIC); Czech Republic -- GACR 24-13049S, CAS LQ100102401,
MEYS LM2023032, CZ.02.1.01/0.0/0.0/16{\textunderscore}013/0001402,
CZ.02.1.01/0.0/0.0/18{\textunderscore}046/0016010 and
CZ.02.1.01/0.0/0.0/17{\textunderscore}049/0008422 and CZ.02.01.01/00/22{\textunderscore}008/0004632;
France -- Centre de Calcul IN2P3/CNRS; Centre National de la Recherche
Scientifique (CNRS); Conseil R\'egional Ile-de-France; D\'epartement
Physique Nucl\'eaire et Corpusculaire (PNC-IN2P3/CNRS); D\'epartement
Sciences de l'Univers (SDU-INSU/CNRS); Institut Lagrange de Paris (ILP)
Grant No.~LABEX ANR-10-LABX-63 within the Investissements d'Avenir
Programme Grant No.~ANR-11-IDEX-0004-02; Germany -- Bundesministerium
f\"ur Bildung und Forschung (BMBF); Deutsche Forschungsgemeinschaft (DFG);
Finanzministerium Baden-W\"urttemberg; Helmholtz Alliance for
Astroparticle Physics (HAP); Helmholtz-Gemeinschaft Deutscher
Forschungszentren (HGF); Ministerium f\"ur Kultur und Wissenschaft des
Landes Nordrhein-Westfalen; Ministerium f\"ur Wissenschaft, Forschung und
Kunst des Landes Baden-W\"urttemberg; Italy -- Istituto Nazionale di
Fisica Nucleare (INFN); Istituto Nazionale di Astrofisica (INAF);
Ministero dell'Universit\`a e della Ricerca (MUR); CETEMPS Center of
Excellence; Ministero degli Affari Esteri (MAE), ICSC Centro Nazionale
di Ricerca in High Performance Computing, Big Data and Quantum
Computing, funded by European Union NextGenerationEU, reference code
CN{\textunderscore}00000013; M\'exico -- Consejo Nacional de Ciencia y Tecnolog\'\i{}a
(CONACYT) No.~167733; Universidad Nacional Aut\'onoma de M\'exico (UNAM);
PAPIIT DGAPA-UNAM; The Netherlands -- Ministry of Education, Culture and
Science; Netherlands Organisation for Scientific Research (NWO); Dutch
national e-infrastructure with the support of SURF Cooperative; Poland
-- Ministry of Education and Science, grants No.~DIR/WK/2018/11 and
2022/WK/12; National Science Centre, grants No.~2016/22/M/ST9/00198,
2016/23/B/ST9/01635, 2020/39/B/ST9/01398, and 2022/45/B/ST9/02163;
Portugal -- Portuguese national funds and FEDER funds within Programa
Operacional Factores de Competitividade through Funda\c{c}\~ao para a Ci\^encia
e a Tecnologia (COMPETE); Romania -- Ministry of Research, Innovation
and Digitization, CNCS-UEFISCDI, contract no.~30N/2023 under Romanian
National Core Program LAPLAS VII, grant no.~PN 23 21 01 02 and project
number PN-III-P1-1.1-TE-2021-0924/TE57/2022, within PNCDI III; Slovenia
-- Slovenian Research Agency, grants P1-0031, P1-0385, I0-0033, N1-0111;
Spain -- Ministerio de Ciencia e Innovaci\'on/Agencia Estatal de
Investigaci\'on (PID2019-105544GB-I00, PID2022-140510NB-I00 and
RYC2019-027017-I), Xunta de Galicia (CIGUS Network of Research Centers,
Consolidaci\'on 2021 GRC GI-2033, ED431C-2021/22 and ED431F-2022/15),
Junta de Andaluc\'\i{}a (SOMM17/6104/UGR and P18-FR-4314), and the European
Union (Marie Sklodowska-Curie 101065027 and ERDF); USA -- Department of
Energy, Contracts No.~DE-AC02-07CH11359, No.~DE-FR02-04ER41300,
No.~DE-FG02-99ER41107 and No.~DE-SC0011689; National Science Foundation,
Grant No.~0450696; The Grainger Foundation; Marie Curie-IRSES/EPLANET;
European Particle Physics Latin American Network; and UNESCO.
\end{sloppypar}

\begin{table}[t]
    \centering
    \caption{Same as \autoref{tab:results}, but using a fixed window position~$(L,B)=(162\degr,-6\degr)$ and separate energy bins \label{tab:disjoint}}
    \begin{tabular}{CCC|hhhCCCCh|hhhhhhhh}
        E_{\min} & E_{\max} & \ntot & 
        L & B & \tfrac\ein\etot & \nbg & \nin & \tfrac\fin\fout & \ZLM & ^{99\%}_\mathrm{U.L.} & 
        L & B & \tfrac\ein\etot & \nbg & \nin & \tfrac\fin\fout & \ZLM & ^{99\%}_\mathrm{U.L.} \\[\smallskipamount]
        \tableline
        20~\EeV & 25~\EeV & 3452 & 162\degr & -6\degr & 9.56\% & 324. & 389 & 1.20^{+0.07}_{-0.06} & +3.3\sigma & 1.36 & 241\degr & \phn{-}5\degr & 10.27\% & 350. & 392 & 1.12^{+0.06}_{-0.06} & +2.1\sigma & 1.27 \\
25~\EeV & 32~\EeV & 2444 & 162\degr & -6\degr & 9.56\% & 233. & 242 & 1.04^{+0.07}_{-0.07} & +0.6\sigma & 1.22 & 320\degr & -19\degr & \phn3.77\% & 91.2 & 114 & 1.25^{+0.12}_{-0.12} & +2.2\sigma & 1.56 \\
32~\EeV & 40~\EeV & 1403 & 162\degr & -6\degr & 9.56\% & 132. & 151 & 1.14^{+0.10}_{-0.10} & +1.5\sigma & 1.39 & \phn75\degr & -27\degr & \phn2.57\% & 35.6 & \phn52 & 1.46^{+0.22}_{-0.20} & +2.5\sigma & 2.02 \\
40~\EeV & 50~\EeV & \phn820 & 162\degr & -6\degr & 9.56\% & 75.4 & 106 & 1.41^{+0.15}_{-0.14} & +3.1\sigma & 1.79 & \phn94\degr & \phn{-}9\degr & \phn3.47\% & 27.7 & \phn50 & 1.81^{+0.28}_{-0.25} & +3.7\sigma & 2.53 \\
50~\EeV & 63~\EeV & \phn418 & 162\degr & -6\degr & 9.56\% & 37.9 & \phn59 & 1.56^{+0.23}_{-0.21} & +3.0\sigma & 2.15 & 331\degr & -26\degr & \phn3.09\% & 12.6 & \phn22 & 1.74^{+0.41}_{-0.35} & +2.3\sigma & 2.86 \\
63~\EeV & +\infty & \phn295 & 162\degr & -6\degr & 9.56\% & 26.6 & \phn43 & 1.62^{+0.28}_{-0.25} & +2.7\sigma & 2.37 & 223\degr & +26\degr & \phn9.56\% & 26.7 & \phn42 & 1.57^{+0.28}_{-0.25} & +2.6\sigma & 2.31 \\

        \tableline
    \end{tabular}
\end{table}

\appendix 

\section{Results using separate energy bins}
\label{app:bins}

\begin{figure*}[h]
    \centering
    \includegraphics{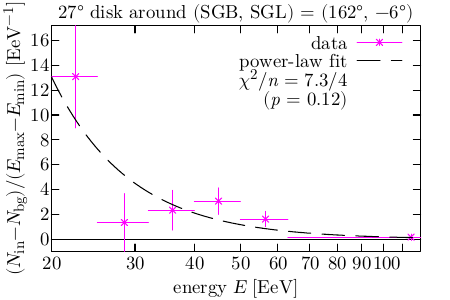}
    \caption{The number of excess events, $\nin-\nbg$, fitted as a power law spectrum $N(E_{\min},E_{\max}) = \int_{E_{\min}}^{E_{\max}} J_{20}\left(\frac{E}{20\,\EeV}\right)^{-\gamma} \dd{E} = N_{20}\frac
        {E_{\min}^{1-\gamma} - E_{\max}^{1-\gamma}}
        {(20\,\EeV)^{1-\gamma}}
    $. The uncertainty on each entry is the sum in quadrature of those on~$\nin$ and~$\nbg$, computed as~$\sqrt{\nin}$ and~$\sqrt{\nout}\ein/\eout$ respectively. In the last bin, we use~$E_{\max} = 166\,\EeV$, the energy of the most energetic event.  The best-fit parameter values we obtain are~$N_{20} = 160 \pm 32$ and~$\gamma = 2.63\pm0.35$.
    Unlike in \citet{spectrum2020}, in this work we do not correct for energy resolution effects; we estimate that here their effect on the spectral index would be an order of magnitude less than the statistical uncertainty of the fit. 
    \label{fig:fit}}
\end{figure*}

In order to describe the energy dependence of the excess in the Centaurus region, we computed~$\nin$, $\nbg$ and~$\ZLM$ in separate bins~$[20\,\EeV, 25\,\EeV)$, $\ldots$, $[50\,\EeV, 63\,\EeV)$, $[63\,\EeV, +\infty)$ rather than cumulative ones~$[20\,\EeV, +\infty)$, $[25\,\EeV, +\infty)$, $\ldots$, keeping the window position fixed to the maximum-significance one found in $[20\,\EeV, +\infty)$.
The results are listed in \autoref{tab:disjoint}.
Also, we fitted a power-law energy spectrum~$\frac{\dd{N}}{\dd{E}} \propto E^{-\gamma}$ integrated over the bins to the excess $\nin-\nbg$, as shown in \autoref{fig:fit}.
While the excess is considerably weaker in the third bin and barely present in the second bin, the behavior is still consistent with a simple power law with the current amount of statistics.

\section{Upper limits as a function of the window position}
\label{app:ul}

\begin{figure*}
    \plotone{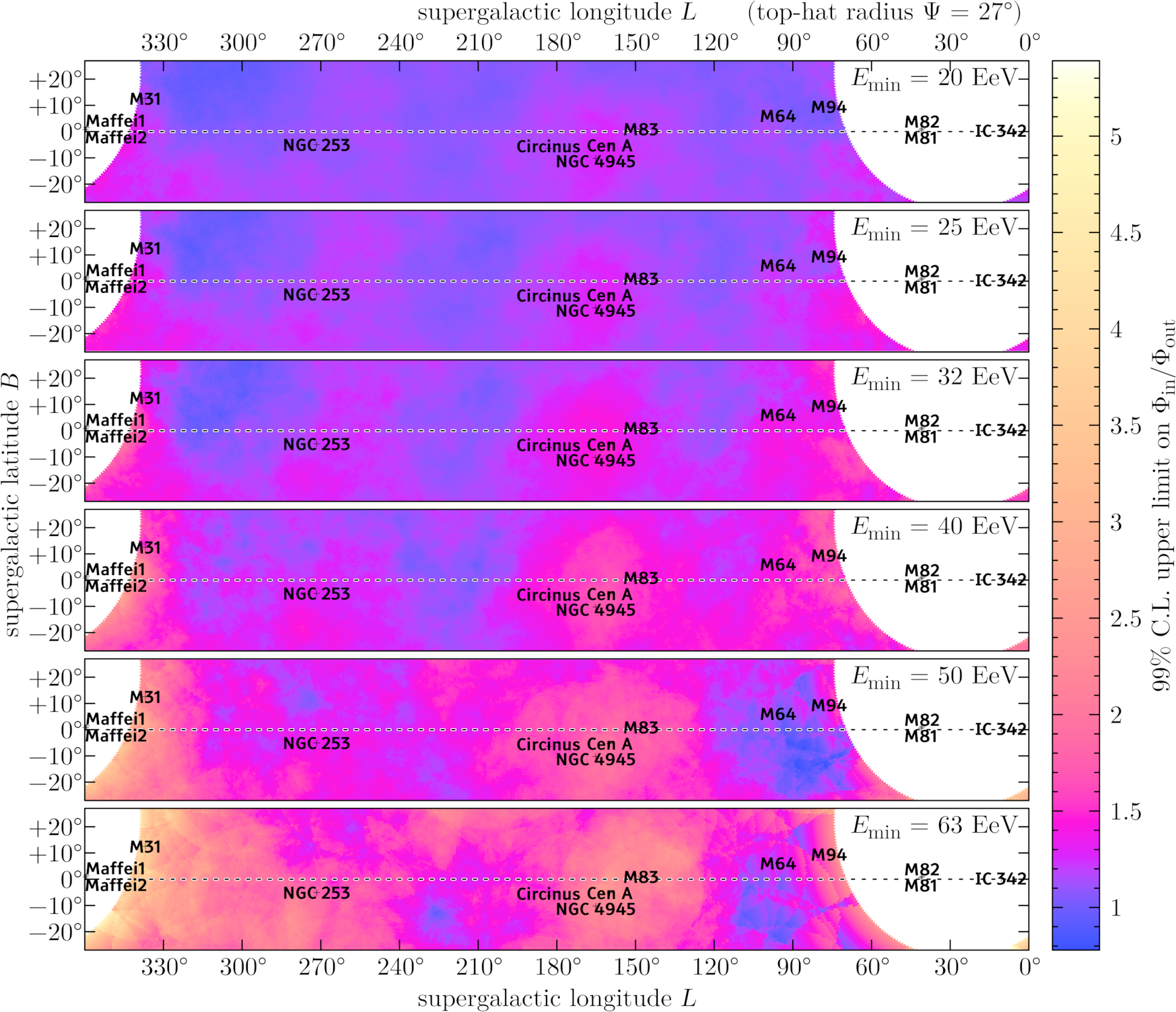}
    \caption{Same as \autoref{fig:lima}, showing the frequentist 99\%~CL upper limit to~$\fin/\fout$ \eqref{eq:upper}.\label{fig:ul}}
\end{figure*}

In \autoref{fig:ul},
we show the frequentist 99\%~CL upper limit to~$\fin/\fout$, as determined from \autoref{eq:upper} for each of the energy thresholds and window positions we considered, showing how our data can set stringent upper limits to the flux in the windows except very close to the edge of the FoV.

\newpage


\begin{thebibliography}{dummy}
    \bibitem[Allard(2012)]{propagation} Allard, D. \journal{\aph}{39}{2012}{33}{10.1016/j.astropartphys.2011.10.011}
    \bibitem[Alves~Batista \etal(2017)]{IGMF} Alves~Batista, R., Shin, M.-S., Devriendt, J., Semikoz, D., \& Sigl, G. \journal{\prd}{96}{2017}{023010}{10.1103/PhysRevD.96.023010}
    \bibitem[Anchordoqui(2023)]{Anchordoqui} Anchordoqui, L. A. \journal{\prd}{107}{2023}{083024}{10.1103/PhysRevD.107.083024}
    \bibitem[GCOS Collaboration(2023)]{GCOS} GCOS Collaboration \journal{\pos}{444}{2023}{281}{10.22323/1.444.0281}, in Proc.\ 38th ICRC, 26 Jul.--3 Aug.\ 2023, Nagoya, Japan
    \bibitem[Górski \etal(2005)]{HEALPix} Górski, K. M., Hivon, E., Banday, A. J., Wandelt, B. D., Hansen, F. K., \etal\ \journal{\apj}{622}{2005}{759}{10.1086/427976}
    \bibitem[GRAND Collaboration(2020)]{GRAND} GRAND Collaboration \journal{\scpma}{63}{2020}{219501}{10.1007/s11433-018-9385-7}
    \bibitem[Li \& Ma(1983)]{LiMa} Li, T.-P., \& Ma, Y.-Q. \journal{\apj}{272}{1983}{317}{10.1086/161295}
    \bibitem[McCall(2014)]{McCall} McCall, M. L. \journal{\mnras}{440}{2014}{405}{10.1093/mnras/stu199}
    \bibitem[Neronov \etal(2023)]{Neronov} Neronov, A., Semikoz, D., \& Kalashev, O. \journal{\prd}{108}{2023}{103008}{10.1103/PhysRevD.108.103008}
    \bibitem[Pierre Auger Collaboration(2010)]{Auger_APh2010} Pierre Auger Collaboration \journal{\aph}{34}{2010}{314}{10.1016/j.astropartphys.2010.08.010}
    \bibitem[Pierre Auger Collaboration(2016)]{Auger_Prime} Pierre Auger Collaboration 2016, \arXiv{1604.03637}
    \bibitem[Pierre Auger Collaboration(2017)]{dipole2017} Pierre Auger Collaboration \journal{\sci}{357}{2017}{1266}{10.1126/science.aan4338}
    \bibitem[Pierre Auger Collaboration(2018a)]{dipole2018} Pierre Auger Collaboration \journal{\apj}{868}{2018a}{4}{10.3847/1538-4357/aae689}
    \bibitem[Pierre Auger Collaboration(2018b)]{Auger_ApJL2018} Pierre Auger Collaboration \journal{\apjl}{853}{2018b}{L29}{10.3847/2041-8213/aaa66d}
    \bibitem[Pierre Auger Collaboration(2020a)]{dipole2020} Pierre Auger Collaboration \journal{\apj}{891}{2020a}{142}{10.3847/1538-4357/ab7236}
    \bibitem[Pierre Auger Collaboration(2020b)]{spectrum2020} Pierre Auger Collaboration \journal{\prd}{102}{2020b}{062005}{10.1103/PhysRevD.102.062005}
    \bibitem[Pierre Auger Collaboration(2021a)]{ML_muons} Pierre Auger Collaboration \journal{\jinst}{16}{2021a}{P07016}{10.1088/1748-0221/16/07/P07016}
    \bibitem[Pierre Auger Collaboration(2021b)]{ML_depth} Pierre Auger Collaboration \journal{\jinst}{16}{2021b}{P07019}{10.1088/1748-0221/16/07/P07019}
    \bibitem[Pierre Auger Collaboration(2022)]{Auger_ApJ2022} Pierre Auger Collaboration \journal{\apj}{935}{2022}{170}{10.3847/1538-4357/ac7d4e} (data and code available at \url{https://doi.org/10.5281/zenodo.6504276})
    \bibitem[Pierre Auger Collaboration(2023)]{Auger_ICRC2023} Pierre Auger Collaboration \journal{\pos}{444}{2023}{252}{10.22323/1.444.0252}, in Proc.\@ 38th ICRC\@, 26 Jul.\@--3 Aug.\@ 2023, Nagoya, Japan
    \bibitem[Pierre Auger Collaboration \& Telescope Array Collaboration(2021)]{joint_ICRC2021} Pierre Auger Collaboration \& Telescope Array Collaboration \journal{\pos}{395}{2021}{308}{10.22323/1.395.0308}, in Proc.\@ 37th ICRC\@, 12--23 Jul.\@ 2021, Berlin, Germany 
    \bibitem[Pierre Auger Collaboration \& Telescope Array Collaboration(2023a)]{joint_UHECR2022} Pierre Auger Collaboration \& Telescope Array Collaboration \journal{\epjwc}{283}{2023a}{03002}{10.1051/epjconf/202328303002}, in Proc.\@ 6th UHECR\@, 3--7 Oct.\@ 2022, L'Aquila, Italy
    \bibitem[Pierre Auger Collaboration \& Telescope Array Collaboration(2023b)]{joint_ICRC2023} Pierre Auger Collaboration \& Telescope Array Collaboration \journal{\pos}{444}{2023b}{521}{10.22323/1.444.0521}, in Proc.\@ 38th ICRC\@, 26 Jul.\@--3 Aug.\@ 2023, Nagoya, Japan
    \bibitem[Plotko \etal(2023)]{Plotko} Plotko, P., van~Vliet, A., Rodrigues, X., \& Winter, W. \journal{\apj}{953}{2023}{129}{10.3847/1538-4357/acdf59}
    \bibitem[POEMMA Collaboration(2021)]{POEMMA} POEMMA Collaboration \journal{\jcap}{06}{2021}{007}{10.1088/1475-7516/2021/06/007}
    \bibitem[Pshirkov \etal(2013)]{turGMF} Pshirkov, M. S., Tinyakov, P. G., \& Urban, F. U. \journal{\mnras}{436}{2013}{2326}{10.1093/mnras/stt1731}
    \bibitem[Sommers(2001)]{Sommers} Sommers, P. \journal{\aph}{14}{2001}{271}{10.1016/S0927-6505(00)00130-4}
    \bibitem[Telescope Array Collaboration(2014)]{TA_ApJL2014} Telescope Array Collaboration \journal{\apjl}{790}{2014}{L21}{10.1088/2041-8205/790/2/L21}
    \bibitem[Telescope Array Collaboration(2018a)]{TA_coldspot} Telescope Array Collaboration \journal{\apj}{862}{2018a}{91}{10.3847/1538-4357/aac9c8}
    \bibitem[Telescope Array Collaboration(2018b)]{TA_decldep} Telescope Array Collaboration 2018b, \arXiv{1801.07820}
    \bibitem[Telescope Array Collaboration(2019)]{TA_BDT} Telescope Array Collaboration \journal{\prd}{99}{2019}{022002}{10.1103/PhysRevD.99.022002}
    \bibitem[Telescope Array Collaboration(2021a)]{TAx4} Telescope Array Collaboration \journal{\nimpa}{1019}{2021a}{165726}{10.1016/j.nima.2021.165726}
    \bibitem[Telescope Array Collaboration(2021b)]{TA_arXiv2021} Telescope Array Collaboration 2021b, \arXiv{2110.14827}
    \bibitem[Telescope Array Collaboration(2023)]{TA_ICRC2023} Telescope Array Collaboration \journal{\pos}{444}{2023}{244}{10.22323/1.444.0244}, in Proc.\@ 38th ICRC\@, 26 Jul.\@--3 Aug.\@ 2023, Nagoya, Japan
    \bibitem[Telescope Array Collaboration(2024)]{TA_arXiv2024} Telescope Array Collaboration 2024, \arXiv{2406.08612}
    \bibitem[Unger \& Farrar(2023)]{cohGMF} Unger, M., \& Farrar, G. R. 2023, \arXiv{2311.12120}
\end{thebibliography}
\newcommand{\journal}[5]{#3, \href{https://doi.org/#5}{#1, #2, #4}}
\newcommand{\arXiv}[1]{\href{https://arxiv.org/abs/#1}{arXiv:#1}}
\newcommand{\etal}{et~al.\@}
\providecommand{\aph}{APh} 
\providecommand{\epjwc}{EPJ Web Conf.} 
\providecommand{\jetpl}{JETPL} 
\providecommand{\jinst}{JInst} 
\providecommand{\nimpa}{NIMPA} 
\providecommand{\pos}{PoS} 
\providecommand{\sci}{Sci} 
\providecommand{\scpma}{SCPMA} 

\newpage

\AuthorCollaborationLimit=3000
{\bf\Large{The Pierre Auger Collaboration}}
\\
A.~Abdul Halim$^{13}$,
P.~Abreu$^{73}$,
M.~Aglietta$^{55,53}$,
I.~Allekotte$^{1}$,
K.~Almeida Cheminant$^{71}$,
A.~Almela$^{7,12}$,
R.~Aloisio$^{46,47}$,
J.~Alvarez-Mu\~niz$^{79}$,
J.~Ammerman Yebra$^{79}$,
G.A.~Anastasi$^{55,53}$
% ...

\renewcommand*\descriptionlabel[1]{\hspace\labelsep\normalfont #1}
\begin{description}[labelsep=0.2em,align=right,labelwidth=0.7em,labelindent=0em,leftmargin=2em,noitemsep]
\item[$^{1}$] Centro At\'omico Bariloche and Instituto Balseiro (CNEA-UNCuyo-CONICET), San Carlos de Bariloche, Argentina
\item[$^{2}$] Departamento de F\'\i{}sica and Departamento de Ciencias de la Atm\'osfera y los Oc\'eanos, FCEyN, Universidad de Buenos Aires and CONICET, Buenos Aires, Argentina
\item[$^{3}$] IFLP, Universidad Nacional de La Plata and CONICET, La Plata, Argentina
\item[$^{4}$] Instituto de Astronom\'\i{}a y F\'\i{}sica del Espacio (IAFE, CONICET-UBA), Buenos Aires, Argentina
\item[$^{5}$] Instituto de F\'\i{}sica de Rosario (IFIR) -- CONICET/U.N.R.\ and Facultad de Ciencias Bioqu\'\i{}micas y Farmac\'euticas U.N.R., Rosario, Argentina
\item[$^{6}$] Instituto de Tecnolog\'\i{}as en Detecci\'on y Astropart\'\i{}culas (CNEA, CONICET, UNSAM), and Universidad Tecnol\'ogica Nacional -- Facultad Regional Mendoza (CONICET/CNEA), Mendoza, Argentina
\item[$^{7}$] Instituto de Tecnolog\'\i{}as en Detecci\'on y Astropart\'\i{}culas (CNEA, CONICET, UNSAM), Buenos Aires, Argentina
\item[$^{8}$] International Center of Advanced Studies and Instituto de Ciencias F\'\i{}sicas, ECyT-UNSAM and CONICET, Campus Miguelete -- San Mart\'\i{}n, Buenos Aires, Argentina
\item[$^{9}$] Laboratorio Atm\'osfera -- Departamento de Investigaciones en L\'aseres y sus Aplicaciones -- UNIDEF (CITEDEF-CONICET), Argentina
\item[$^{10}$] Observatorio Pierre Auger, Malarg\"ue, Argentina
\item[$^{11}$] Observatorio Pierre Auger and Comisi\'on Nacional de Energ\'\i{}a At\'omica, Malarg\"ue, Argentina
\item[$^{12}$] Universidad Tecnol\'ogica Nacional -- Facultad Regional Buenos Aires, Buenos Aires, Argentina
\item[$^{13}$] University of Adelaide, Adelaide, S.A., Australia
\item[$^{14}$] Universit\'e Libre de Bruxelles (ULB), Brussels, Belgium
\item[$^{15}$] Vrije Universiteit Brussels, Brussels, Belgium
\item[$^{16}$] Centro Brasileiro de Pesquisas Fisicas, Rio de Janeiro, RJ, Brazil
\item[$^{17}$] Centro Federal de Educa\c{c}\~ao Tecnol\'ogica Celso Suckow da Fonseca, Petropolis, Brazil
\item[$^{18}$] Instituto Federal de Educa\c{c}\~ao, Ci\^encia e Tecnologia do Rio de Janeiro (IFRJ), Brazil
\item[$^{19}$] Universidade de S\~ao Paulo, Escola de Engenharia de Lorena, Lorena, SP, Brazil
\item[$^{20}$] Universidade de S\~ao Paulo, Instituto de F\'\i{}sica de S\~ao Carlos, S\~ao Carlos, SP, Brazil
\item[$^{21}$] Universidade de S\~ao Paulo, Instituto de F\'\i{}sica, S\~ao Paulo, SP, Brazil
\item[$^{22}$] Universidade Estadual de Campinas, IFGW, Campinas, SP, Brazil
\item[$^{23}$] Universidade Estadual de Feira de Santana, Feira de Santana, Brazil
\item[$^{24}$] Universidade Federal de Campina Grande, Centro de Ciencias e Tecnologia, Campina Grande, Brazil
\item[$^{25}$] Universidade Federal do ABC, Santo Andr\'e, SP, Brazil
\item[$^{26}$] Universidade Federal do Paran\'a, Setor Palotina, Palotina, Brazil
\item[$^{27}$] Universidade Federal do Rio de Janeiro, Instituto de F\'\i{}sica, Rio de Janeiro, RJ, Brazil
\item[$^{28}$] Universidade Federal do Rio de Janeiro (UFRJ), Observat\'orio do Valongo, Rio de Janeiro, RJ, Brazil
\item[$^{29}$] Universidade Federal Fluminense, EEIMVR, Volta Redonda, RJ, Brazil
\item[$^{30}$] Universidad de Medell\'\i{}n, Medell\'\i{}n, Colombia
\item[$^{31}$] Universidad Industrial de Santander, Bucaramanga, Colombia
\item[$^{32}$] Charles University, Faculty of Mathematics and Physics, Institute of Particle and Nuclear Physics, Prague, Czech Republic
\item[$^{33}$] Institute of Physics of the Czech Academy of Sciences, Prague, Czech Republic
\item[$^{34}$] Palacky University, Olomouc, Czech Republic
\item[$^{35}$] CNRS/IN2P3, IJCLab, Universit\'e Paris-Saclay, Orsay, France
\item[$^{36}$] Laboratoire de Physique Nucl\'eaire et de Hautes Energies (LPNHE), Sorbonne Universit\'e, Universit\'e de Paris, CNRS-IN2P3, Paris, France
\item[$^{37}$] Univ.\ Grenoble Alpes, CNRS, Grenoble Institute of Engineering Univ.\ Grenoble Alpes, LPSC-IN2P3, 38000 Grenoble, France
\item[$^{38}$] Universit\'e Paris-Saclay, CNRS/IN2P3, IJCLab, Orsay, France
\item[$^{39}$] Bergische Universit\"at Wuppertal, Department of Physics, Wuppertal, Germany
\item[$^{40}$] Karlsruhe Institute of Technology (KIT), Institute for Experimental Particle Physics, Karlsruhe, Germany
\item[$^{41}$] Karlsruhe Institute of Technology (KIT), Institut f\"ur Prozessdatenverarbeitung und Elektronik, Karlsruhe, Germany
\item[$^{42}$] Karlsruhe Institute of Technology (KIT), Institute for Astroparticle Physics, Karlsruhe, Germany
\item[$^{43}$] RWTH Aachen University, III.\ Physikalisches Institut A, Aachen, Germany
\item[$^{44}$] Universit\"at Hamburg, II.\ Institut f\"ur Theoretische Physik, Hamburg, Germany
\item[$^{45}$] Universit\"at Siegen, Department Physik -- Experimentelle Teilchenphysik, Siegen, Germany
\item[$^{46}$] Gran Sasso Science Institute, L'Aquila, Italy
\item[$^{47}$] INFN Laboratori Nazionali del Gran Sasso, Assergi (L'Aquila), Italy
\item[$^{48}$] INFN, Sezione di Catania, Catania, Italy
\item[$^{49}$] INFN, Sezione di Lecce, Lecce, Italy
\item[$^{50}$] INFN, Sezione di Milano, Milano, Italy
\item[$^{51}$] INFN, Sezione di Napoli, Napoli, Italy
\item[$^{52}$] INFN, Sezione di Roma ``Tor Vergata'', Roma, Italy
\item[$^{53}$] INFN, Sezione di Torino, Torino, Italy
\item[$^{54}$] Istituto di Astrofisica Spaziale e Fisica Cosmica di Palermo (INAF), Palermo, Italy
\item[$^{55}$] Osservatorio Astrofisico di Torino (INAF), Torino, Italy
\item[$^{56}$] Politecnico di Milano, Dipartimento di Scienze e Tecnologie Aerospaziali , Milano, Italy
\item[$^{57}$] Universit\`a del Salento, Dipartimento di Matematica e Fisica ``E.\ De Giorgi'', Lecce, Italy
\item[$^{58}$] Universit\`a dell'Aquila, Dipartimento di Scienze Fisiche e Chimiche, L'Aquila, Italy
\item[$^{59}$] Universit\`a di Catania, Dipartimento di Fisica e Astronomia ``Ettore Majorana``, Catania, Italy
\item[$^{60}$] Universit\`a di Milano, Dipartimento di Fisica, Milano, Italy
\item[$^{61}$] Universit\`a di Napoli ``Federico II'', Dipartimento di Fisica ``Ettore Pancini'', Napoli, Italy
\item[$^{62}$] Universit\`a di Palermo, Dipartimento di Fisica e Chimica ''E.\ Segr\`e'', Palermo, Italy
\item[$^{63}$] Universit\`a di Roma ``Tor Vergata'', Dipartimento di Fisica, Roma, Italy
\item[$^{64}$] Universit\`a Torino, Dipartimento di Fisica, Torino, Italy
\item[$^{65}$] Benem\'erita Universidad Aut\'onoma de Puebla, Puebla, M\'exico
\item[$^{66}$] Unidad Profesional Interdisciplinaria en Ingenier\'\i{}a y Tecnolog\'\i{}as Avanzadas del Instituto Polit\'ecnico Nacional (UPIITA-IPN), M\'exico, D.F., M\'exico
\item[$^{67}$] Universidad Aut\'onoma de Chiapas, Tuxtla Guti\'errez, Chiapas, M\'exico
\item[$^{68}$] Universidad Michoacana de San Nicol\'as de Hidalgo, Morelia, Michoac\'an, M\'exico
\item[$^{69}$] Universidad Nacional Aut\'onoma de M\'exico, M\'exico, D.F., M\'exico
\item[$^{70}$] Universidad Nacional de San Agustin de Arequipa, Facultad de Ciencias Naturales y Formales, Arequipa, Peru
\item[$^{71}$] Institute of Nuclear Physics PAN, Krakow, Poland
\item[$^{72}$] University of \L{}\'od\'z, Faculty of High-Energy Astrophysics,\L{}\'od\'z, Poland
\item[$^{73}$] Laborat\'orio de Instrumenta\c{c}\~ao e F\'\i{}sica Experimental de Part\'\i{}culas -- LIP and Instituto Superior T\'ecnico -- IST, Universidade de Lisboa -- UL, Lisboa, Portugal
\item[$^{74}$] ``Horia Hulubei'' National Institute for Physics and Nuclear Engineering, Bucharest-Magurele, Romania
\item[$^{75}$] Institute of Space Science, Bucharest-Magurele, Romania
\item[$^{76}$] Center for Astrophysics and Cosmology (CAC), University of Nova Gorica, Nova Gorica, Slovenia
\item[$^{77}$] Experimental Particle Physics Department, J.\ Stefan Institute, Ljubljana, Slovenia
\item[$^{78}$] Universidad de Granada and C.A.F.P.E., Granada, Spain
\item[$^{79}$] Instituto Galego de F\'\i{}sica de Altas Enerx\'\i{}as (IGFAE), Universidade de Santiago de Compostela, Santiago de Compostela, Spain
\item[$^{80}$] IMAPP, Radboud University Nijmegen, Nijmegen, The Netherlands
\item[$^{81}$] Nationaal Instituut voor Kernfysica en Hoge Energie Fysica (NIKHEF), Science Park, Amsterdam, The Netherlands
\item[$^{82}$] Stichting Astronomisch Onderzoek in Nederland (ASTRON), Dwingeloo, The Netherlands
\item[$^{83}$] Universiteit van Amsterdam, Faculty of Science, Amsterdam, The Netherlands
\item[$^{84}$] Case Western Reserve University, Cleveland, OH, USA
\item[$^{85}$] Colorado School of Mines, Golden, CO, USA
\item[$^{86}$] Department of Physics and Astronomy, Lehman College, City University of New York, Bronx, NY, USA
\item[$^{87}$] Michigan Technological University, Houghton, MI, USA
\item[$^{88}$] New York University, New York, NY, USA
\item[$^{89}$] University of Chicago, Enrico Fermi Institute, Chicago, IL, USA
\item[$^{90}$] University of Delaware, Department of Physics and Astronomy, Bartol Research Institute, Newark, DE, USA
\item[$^{91}$] University of Wisconsin-Madison, Department of Physics and WIPAC, Madison, WI, USA
\item[] -----
\item[$^{a}$] Louisiana State University, Baton Rouge, LA, USA
\item[$^{b}$] Institut universitaire de France (IUF), France
\item[$^{c}$] also at University of Bucharest, Physics Department, Bucharest, Romania
\item[$^{d}$] School of Physics and Astronomy, University of Leeds, Leeds, United Kingdom
\item[$^{e}$] now at Agenzia Spaziale Italiana (ASI).\ Via del Politecnico 00133, Roma, Italy
\item[$^{f}$] Fermi National Accelerator Laboratory, Fermilab, Batavia, IL, USA
\item[$^{g}$] now at Graduate School of Science, Osaka Metropolitan University, Osaka, Japan
\item[$^{h}$] now at ECAP, Erlangen, Germany
\item[$^{i}$] Max-Planck-Institut f\"ur Radioastronomie, Bonn, Germany
\item[$^{j}$] also at Kapteyn Institute, University of Groningen, Groningen, The Netherlands
\item[$^{k}$] Colorado State University, Fort Collins, CO, USA
\item[$^{l}$] Pennsylvania State University, University Park, PA, USA
\end{description}


\end{document}